# X-Ray Spectral Shape Variation in Changing-look Seyfert Galaxy SDSS J155258+273728

Yanli Ai[1], Liming Dou[2,3], Chenwei Yang[4,5], Luming Sun[5,6], Fu-Guo Xie[7], Su Yao[8], Xue-Bing Wu[8,9], Tinggui Wang[5,6], Xinwen Shu[10], and Ning Jiang[5,6]
[1] School of Physics and Astronomy, Sun Yat-Sen University, Guangzhou 510275, People's Republic of China; aiyanli@mail.sysu.edu.cn
[2] Center for Astrophysics, Guangzhou University, Guangzhou 510006, People's Republic of China
[3] Astronomy Science and Technology Research Laboratory of Department of Education of Guangdong Province, Guangzhou 510006, People's Republic of China
[4] Antarctic Astronomy Research Division, Key Laboratory for Polar Science of the State Oceanic Administration, Polar Research Institute of China, Shanghai, People's Republic of China
[5] School of Astronomy and Space Sciences, University of Science and Technology of China, Hefei, People's Republic of China
[6] Key Laboratory for Research in Galaxies and Cosmology of Chinese Academy of Sciences, Department of Astronomy, University of Science and Technology of China, Hefei, People's Republic of China
[7] Key Laboratory for Research in Galaxies and Cosmology, Shanghai Astronomical Observatory, Chinese Academy of Sciences, 80 Nandan Road, Shanghai 200030, People's Republic of China
[8] Kavli Institute for Astronomy and Astrophysics, Peking University, Beijing 100871, People's Republic of China
[9] Department of Astronomy, School of Physics, Peking University, Beijing 100871, People's Republic of China
[10] Department of Physics, Anhui Normal University, Wuhu, Anhui, 241000, People's Republic of China
Received 2019 November 28; revised 2020 February 1; accepted 2020 February 5; published 2020 February 25

## Abstract

We analyze the X-ray, optical, and mid-infrared data of a "changing-look" Seyfert galaxy SDSS J155258+273728 at $z \simeq 0.086$. Over a period of one decade (2009–2018), its broad H$\alpha$ line intensity increased by a factor of $\sim$4. Meanwhile, the X-ray emission in 2014 as observed by *Chandra* was about five times brighter than that in 2010 by *Suzaku*, and the corresponding emissions in the *V*-band, mid-infrared *W*1 band brighten by $\sim$0.18, 0.32 mag, respectively. Moreover, the absorption in X-rays is moderate and stable, i.e., $N_{\rm H} \sim 10^{21}$ cm$^{-2}$, but the X-ray spectrum becomes harder in the 2014 *Chandra* bright state (i.e., photon index $\Gamma = 1.52^{+0.06}_{-0.06}$) than that of the 2010 *Suzaku* low state ($\Gamma = 2.03^{+0.22}_{-0.21}$). With an Eddington ratio being lower than a few percent, the inner region of the accretion disk in SDSS J155258+273728 is likely a hot accretion flow. We then compile from literature the X-ray data of "changing-look" active galactic nuclei (AGNs), and find that they generally follow the well-established "*V*"-shaped correlation in AGNs, that is, above a critical turnover luminosity the X-ray spectra soften with the increasing luminosity, and below that luminosity the trend is reversed in the way of "harder when brighter." This presents direct evidence that CL-AGNs have distinctive changes in not only the optical spectral type, but also the X-ray spectral shape. The similarity in the X-ray spectral evolution between CL-AGNs and black hole X-ray binaries indicates that the observed CL-AGNs phenomena may relate to the state transition in accretion physics.

*Unified Astronomy Thesaurus concepts:* Galaxies (573); Active galactic nuclei (16); Quasars (1319)

## 1. Introduction

The optical/UV continuum variability of active galactic nuclei (AGNs)/quasars on timescales of months to years with typical fluctuations of $\simeq$10% has long been known and was well studied in recent years with wide-area, multi-epoch optical surveys (e.g., Vanden Berk et al. 2004; Ai et al. 2013; Li et al. 2018; Sun et al. 2018). This continuous and stochastic variability is mostly explained in the context of standard accretion disk theory with different scenarios, i.e., reprocessing of emission from near the black hole (e.g., McHardy et al. 2014; Edelson et al. 2019), local instabilities on a thermal timescale (e.g., Kelly et al. 2009; Cai et al. 2016, 2018), or changes of global mass accretion rate with confinement of a viscous timescale (e.g., Liu et al. 2016). Well-sampled spectroscopic monitoring has shown that the variability in broad emission-line fluxes well correlated with that of the continuum as "light echoes," and the delay measured with "reverberation mapping" technique is used to probe the structure and kinematics of the line-emitting gas (e.g., Kaspi et al. 2005; Peterson 2014; Shen et al. 2016; Du et al. 2018).

Extreme spectroscopic and photometric variability is detected in a small fraction of AGNs, which are intriguing for investigation of changes of accretion states in proximity to black hole and circumnuclear gas. Large optical luminosity variations by factor of $\gtrsim$2 were detected in quasars (e.g., Lawrence et al. 2016; Rumbaugh et al. 2018; Graham et al. 2020). Disappearing or appearing broad Balmer emission lines have been known for many years in a number of local low-luminosity AGNs (e.g., Tohline & Osterbrock 1976; Cohen et al. 1986; Denney et al. 2014; Shappee et al. 2014). With large-scale time-domain survey recent studies have identified a growing number of such cases in quasars with associated large, order-of-magnitude variations in the optical continuum on month to year timescales (LaMassa et al. 2015; MacLeod et al. 2016; Ruan et al. 2016; Runco et al. 2016; Runnoe et al. 2016; Stern et al. 2018; Wang et al. 2018; Yang et al. 2018; MacLeod et al. 2019; Trakhtenbrot et al. 2019; Sheng et al. 2020). Their optical classification was caught to change between type 1.8–2 (narrow-line) to type 1 (broad-line) AGNs (or vice versa), and even from low-ionization nuclear emission-line region galaxies to broad-line quasars (Gezari et al. 2017; Frederick et al. 2019).

These objects were known as "changing-look active galactic nuclei" (CL-AGNs). The first case with dramatic diminishment of Mg II in one CL-AGN was reported in Guo et al. (2019).

Recent evidence from spectropolarimetry (e.g., Hutsemékers et al. 2019) and mid-infrared (mid-IR) echo (Sheng et al. 2017) implies that variable obscuration is a rather unlikely explanation for the CL-AGNs. The drastic spectral changes seen in CL-AGNs most plausibly come from the intrinsic changes in accretion power. The timescales, from months to years, relevant for the CL-AGNs reported up until now, are far shorter than what is expected for global accretion rate changes in the standard think disk (e.g., Stern et al. 2018). Different models are proposed to address the timescale problem, while all clearly make out that dramatic variations take place at the inner radius of accretion gas are responsible for the changing-look phenomena, either from disk instability or a switch in the nature of accretion flow, or both (Noda & Done 2018; Ross et al. 2018; Stern et al. 2018; Dexter et al. 2019b; Dexter & Begelman 2019a; Śniegowska & Czerny 2019).

Recently, a link between the state transition in black hole X-ray binaries (BHBs) and the observations in CL-AGNs is accumulating. For example, the UV-to-X-ray spectral index and Eddington ratio of CL-AGNs follows a correlation that is similar to the spectral evolution of BHBs (Ruan et al. 2019). Another example comes from two CL-AGNs, Mrk 590 and Mrk 1018. The soft X-ray excess in these two systems disappeared as they decayed, similar to the disappearance of thermal emission during the soft-to-hard state transition in BHBs (Rivers et al. 2012; Noda & Done 2018). More multiband monitoring of CL-AGNs, especially in X-rays, will provide clues about the physics process that occurred in the region proximity to black holes.

The Seyfert galaxy, SDSS J155258+273728 at $z \simeq 0.086$, was recently identified as a "turn-on" CL-AGN with a significantly increased broad H$\alpha$ line emission (Yang et al. 2018). This object was classified as Seyfert 1.9 in Osterbrock (1981) with only a weak broad H$\alpha$ emission line. We present mid-IR, optical to X-ray photometric and spectral monitoring data of the source from 2006 to 2018 (Section 2). From the multiband variability analysis we confirm that CL-AGNs are powered by intrinsic variations in accretion power (Section 3). We probe the X-ray spectral evolution for CL-AGNs and provide clear evidence that the X-ray spectral shape changes with the optical spectral types transition (Section 4). Our results indicate that the CL-AGN phenomena are mostly related with or driven by accretion state transitions.

## 2. Observations

We proposed spectroscopic observation with the double spectrograph (DBSP) of the Hale 200-inch telescope and *Swift* observation in 2018. The object was also observed with *Suzaku* in 2010 and *Chandra* in 2014. We summarize the data sets and reduction procedures used here.

### 2.1. Optical Spectra

SDSS J155258+273728 was observed by SDSS on 2005 May 8 through a 3″ diameter fiber over the wavelength range 3800–9200 Å at a spectral resolution of $\simeq 2000$ (Abazajian et al. 2009). The spectrum, with relatively weak emission lines, is dominated by the host galaxy starlight emission (Figure 1). The object was included in the LAMOST quasar survey with spectroscopic observation on 2014 March 5 (Ai et al. 2016; Dong et al. 2018; Yao et al. 2019). LAMOST is a 4 m reflecting Schmidt telescope equipped with 4000 fibers of 3″ diameter each (Zhao et al. 2012). The wavelength coverage ranges from 3700 to 9000 Å with an overall spectral resolution of $\simeq 1800$. The LAMOST spectrum was scaled to match the red spectrum with the blue spectrum. Compared to SDSS, the significant feature in LAMOST spectrum is the prominent broad H$\alpha$ emission line, as shown in Figure 1.

SDSS J155258+273728 was followed up with the optical DBSP of the Hale 200-inch telescope at Palomar Observatory (P200) on 2018 February 24. Observation was taken through a 1″.0 slit width, using a D55 dichroic, a 600/4000 grating for the blue side, and a 316/7500 grating for the red side. The grating angles were adjusted to obtain a nearly continuous wavelength coverage from 3300 to 10000 Å except for a small gap of 5500–5550 Å. The data were reduced following the standard routine. The spectrum was extracted with a 2″ aperture and flux calibrated using the standard star. The significant broad H$\alpha$ line is also clearly shown in the DBSP spectrum (Figure 1).

### 2.2. Optical Spectral Fitting

To compare the strengths of emission lines among different observations, we recalibrated LAMOST and DBSP spectra with that of Sloan Digital Sky Survey (SDSS) assuming the [O III] $\lambda 5007$ line is not variable over a timescale of 10 yr. As shown in Figure 1, the spectra of SDSS J155258+273728 are dominated by host galaxy starlight rather than the emission from the active nucleus itself in all the three epochs. We first fit a starlight model to the SDSS spectrum to constrain the stellar component. Prior to fitting, all spectra were shifted back to the rest frame and corrected for Galactic extinction.

We fit the SDSS spectrum with the BC03 stellar population model (Bruzual & Charlot 2003) using the STARLIGHT code (Cid Fernandes et al. 2005). In the fitting we masked all the prominent emission lines. The resulting starlight model matched the continuum well and no extra nonstellar component is needed (Figure 1). We then scaled the starlight model from SDSS to match the spectra in DBSP and LAMOST with scale factors chosen by minimization of the residuals in the stellar continuum and absorption lines. For both DBSP and LAMOST spectra, there are also no additional power-law components from the AGN continuum required. We then fit the emission lines in the residual spectra.

The Gaussian function is used to model the emission lines, and we generally follow the method in Ai et al. (2016) with only minor modifications. The H$\alpha$ emission line was modeled with one narrow and one broad component. The broad component is fitted with three Gaussians. The velocity offsets and line widths of [N II] $\lambda\lambda 6548, 6584$ and [S II] $\lambda\lambda 6717, 6731$ are tied to those of the H$\alpha$ narrow component. The relative flux ratio of the two [N II] components is fixed to 2.96. The upper limits of the FWHM for the narrow lines are set to be 1200 km s$^{-1}$. No broad component was detected in the H$\beta$ line in all three of the observations. Thus we model the H$\beta$ line with only one narrow component. Each line of the [O III] $\lambda\lambda 4959, 5007$ doublet is modeled with one Gaussian, and the doublets are assumed to have the same redshifts and profiles, with the flux ratio fixed to the theoretical value of 3. Velocity offsets and line widths of the doublet core component are tied to those of the narrow H$\beta$ component. The fitted models, along with individual components, for the H$\alpha$ and H$\beta$ emission lines in different epochs are shown in Figure 2. The parameters are summarized in Table 1. Dramatic change occurs in the broad H$\alpha$ line between 2005 and 2014/2018. The uncertainties quoted for each parameter are calculated through $10^3$ Monte Carlo simulations of each spectrum based on their 1$\sigma$ flux density uncertainties. The spectral fitting procedure is performed for each spectrum, and 1$\sigma$ spreads in the resulting distributions of resampled parameters are reported as uncertainties.

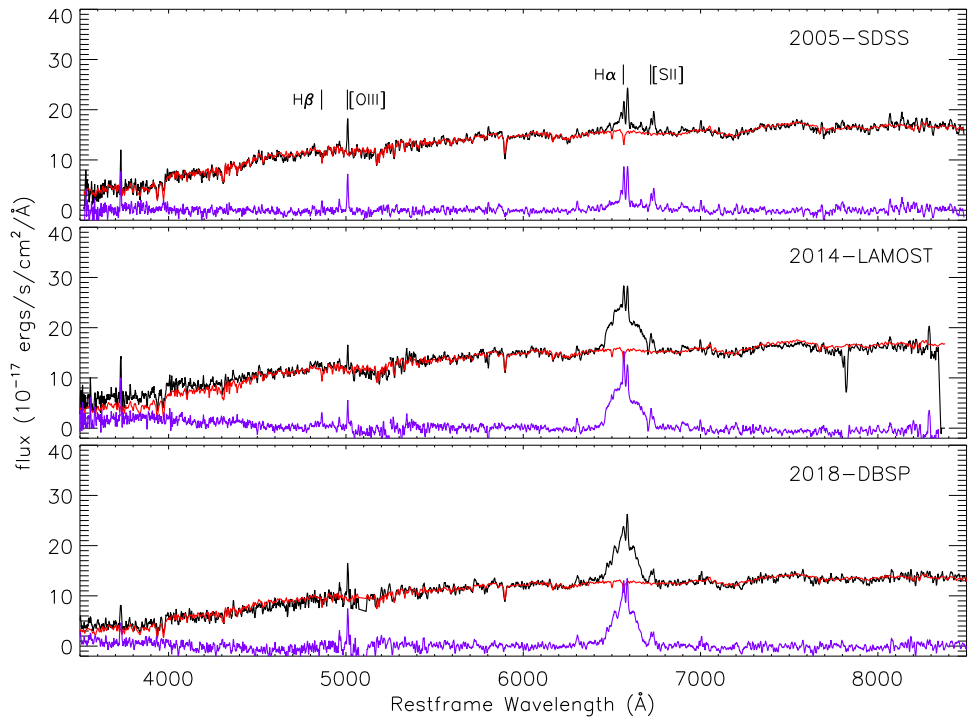

**Figure 1.** Optical spectra of SDSS J155258+273728. The black curves show the observed spectra; the red curves show a host galaxy starlight template that was fit to the SDSS spectrum and scaled to LAMOST and DBSP spectra. The purple curves show the residuals with starlight components subtracted from the observed spectra.

### 2.3. X-Ray Observations and Data Reduction

*Suzaku* observed SDSS J155258+273728 on 2010 July 31 with an exposure of ∼61 ks. We reduced the data with the HEASOFT software package (v6.21), following the procedure outlined in the Suzaku Data Reduction Guide (v5.0). To extract science products from the XIS units, we reprocessed the unfiltered event files for each of the operational CCDs (XIS0, 1, 3) and editing modes (3 × 3, 5 × 5). Cleaned event files were generated by running the *Suzaku* aepipeline pipeline with the latest calibration and screening criteria files (XIS caldb v20160607). The data in the back-illuminated CCD (XIS1) were not used due to the high background level.

SDSS J155258+273728 was also targeted on 2014 December 13 with *Chandra* for ∼84 ks (PI: Kaastra) using the Advanced CCD Imaging Spectrometer (ACIS) instrument. We processed the data with standard CIAO version 4.7 and only events with grades of 0, 2, 3, 4, and 6 were considered in the analysis.

SDSS J155258+273728 was clearly detected in both *Suzaku* and *Chandra* observations. In *Suzaku* XIS images, there are two other peaks identified, each at ∼150″ away from SDSS J155258+273728. These two sources are also detected in the *Chandra* image. With the excellent spatial resolution of *Chandra* ACIS the emissions from these two peaks are well constrained, and both of them are about two orders of magnitude fainter than SDSS J155258+273728. In *Suzaku* observation the X-ray spectral shape extracted from the circle with inclusion of all the three peaks is nearly the same as that extracted from with only inclusion of the peak of SDSS J155258+273728. Thus we conclude that the X-ray emission of SDSS J155258+273728 dominates in the field of the *Suzaku* XIS image. Meanwhile, to get a clean spectrum, we restrict the spectral extraction region to a small circle with a radius of a 80″, corresponding to an encircled energy of ∼60%.[11] The background was extracted from adjacent regions free of any contaminating sources, with care taken to avoid the calibration sources in the corners. Response matrices and ancillary response files were produced with *xisrmfgen* and *xissimarfgen*. Using the FTOOL ADDASCASPEC we combine the spectra and response files for the two front-illuminated detectors (XIS0, 3). The combined spectrum was grouped with at least 10 counts in each energy bin.

For *Chandra* ACIS data we extract the source spectrum from a 3″ circular region. The source spectrum, background spectrum, response matrix files, and auxiliary matrix files are built using the CIAO script SPECEXTRACT. The *Chandra* spectrum was grouped with 15 counts per bin.

We requested and were granted a 2 ks *Swift* X-ray Telescope (XRT) observation on 2018 March 28 for SDSS J155258 +273728. There are three archival XRT observations (ObsID: 00611599000-2) with a total exposure of 24 ks from 2014 September 3 to September 5, of which SDSS J155258 +273728 happened to be in the field of view. The XRT observations were processed with the UK Swift Data Science Centre pipeline, which takes into account dead columns and vignetting to extract counts from the source in the energy range of 0.3–10 keV. We combined the three spectra in 2014 as no significant variability was detected between the observations. The X-ray count rate in 2014 and 2018 was $0.0029 \pm 0.0003$ and $0.0048 \pm 0.0017$ counts s$^{-1}$, respectively. We group the XRT spectra with 10 and 2 counts in each energy bin for the 2014 and 2018 epochs.

---
[11] https://heasarc.gsfc.nasa.gov/docs/suzaku/prop_tools/suzaku_td/node9.html

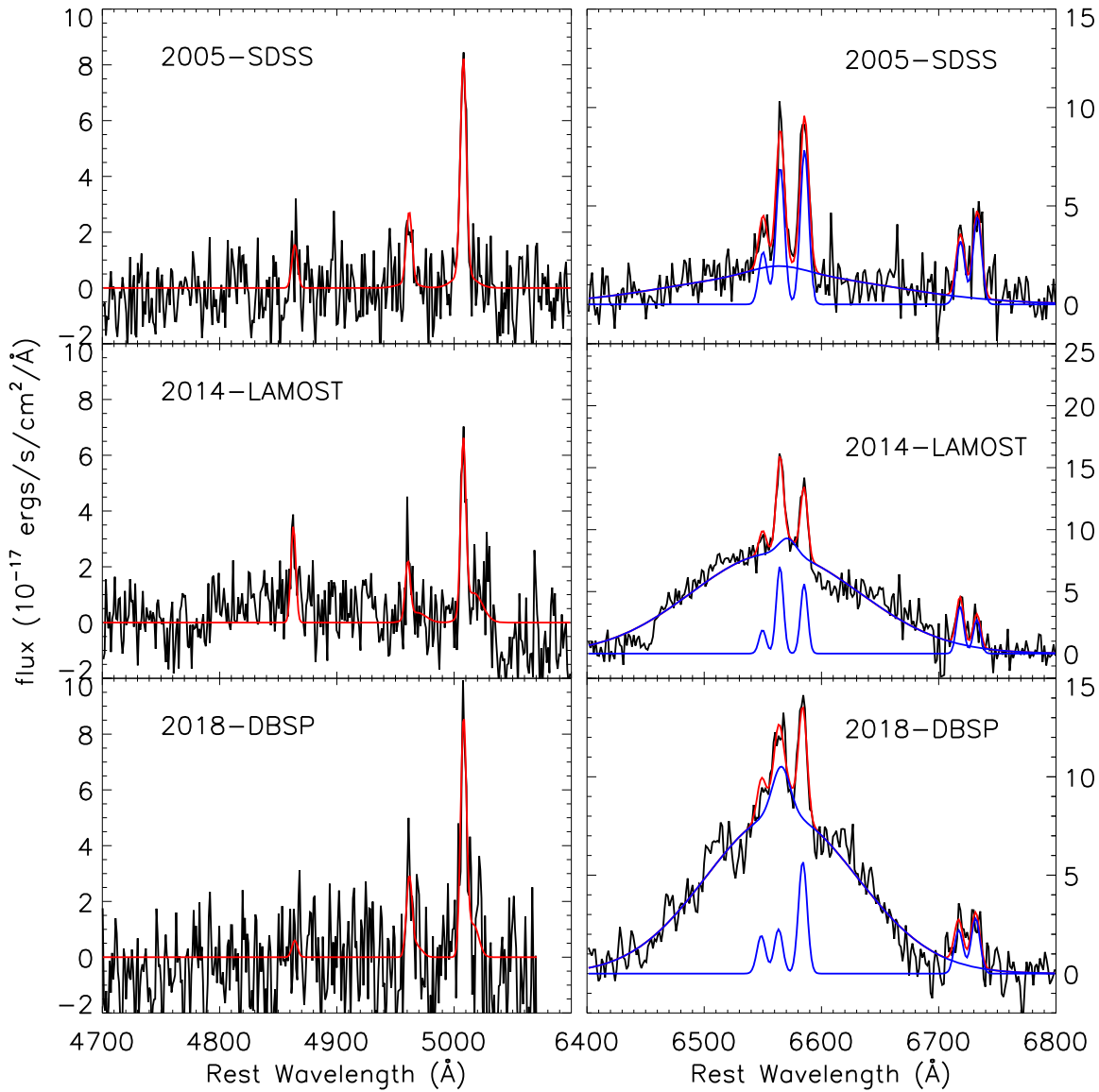

**Figure 2.** Starlight-subtracted spectra of SDSS J155258+273728 zoomed to exhibit the H$\beta$ and H$\alpha$ emission-line regions. The combination of models (red) and individual components (blue) are shown.

## 3. X-Ray Spectroscopy

We fit the X-ray spectra of SDSS J155258+273728 using XSPEC (v12.9, Arnaud 1996). A simple power-law model modified by Galactic absorption with $N_H$ of $3.02 \times 10^{20}$ cm$^{-2}$ (Kalberla et al. 2005) is applied to all spectra at first. The fit is acceptable for the *Suzaku* spectrum, while for the *Chandra* spectrum the residuals at soft energy bands less than 1 keV require additional absorption. We then fold the model with intrinsic absorption (at $z = 0.086$). The improvement is significant with $\Delta\chi^2$ of 33 and the fit is acceptable. We also investigate other features in *Chandra* spectra, which have the highest signal-to-noise ratio (S/N) among the four epochs. There are some residuals at energy ∼7 keV. We then add one Gaussian emission line to fit the feature, while the improvement is not significant. There are no more features that warrant more warm absorption or reflection modeling. We then fit all the spectra with an absorbed power-law model.

The required additional absorption at a redshift of the source in *Chandra* spectra modeling is moderate with a fitted absorption of $N_H = 9.49^{+3.25}_{-3.05} \times 10^{20}$ cm$^{-2}$. The value of the inferred absorption in *Suzaku* is consistent with that within uncertainties (Table 2). The *Chandra* spectra are harder with a photon index of $1.52^{+0.06}_{-0.06}$, compared to the value of $2.03^{+0.22}_{-0.21}$ in *Suzaku*. The two *Swift* spectra are even harder than that of *Chandra*, while the value of the photon index cannot be well constrained due to low S/N and degeneracy with absorption. Thus in the fitting of *Swift* spectra we tied all the parameters to those of *Chandra* except the normalization. The fitted parameters are shown in Table 2, and the spectra with the fitted power-law model are shown in Figure 3. In Figure 3 we also

Table 1
Parameters of Emission-line Fits[a]

|  | 2005–SDSS | | 2014–LAMOST | | 2018–DBSP | |
| --- | --- | --- | --- | --- | --- | --- |
|  | Flux | FWHM | Flux | FWHM | Flux | FWHM |
| | | | Broad Lines | | | |
| H$\alpha$ | 357.9 ± 22.0 | 7736 ± 1992 | 1461.1 ± 53.7 | 6838 ± 897 | 1345.6 ± 34.3 | 5410 ± 642 |
| H$\beta$ | <87.7 | ... | <70.6 | ... | <64.6 | ... |
| | | | Narrow Lines | | | |
| H$\alpha$ | 62.9 ± 5.8 | 386 ± 21 | 53.6 ± 16.8 | 326 ± 59 | 19.7 ± 17.8 | 397 ± 40 |
| H$\beta$ | 9.5 ± 2.8 | 351 ± 50 | 18.1 ± 14.9 | 304 ± 227 | 3.4 ± 5.7 | 350 ± 134 |

**Note.**
[a] Observed flux in units of $10^{-17}$ erg cm$^{-2}$ s$^{-1}$ and FWHM in units of km s$^{-1}$.

Table 2
Parameters of X-Ray Spectra Fits[a]

| Satellite | Date of Obs. | $N_H$ ($10^{20}$ cm$^{-2}$) | $\Gamma$ | $F_{2-10\,\mathrm{keV}}$[a] ($10^{-13}$ erg s$^{-1}$ cm$^{-2}$) |
| --- | --- | --- | --- | --- |
| Suzaku | 2010.07 | $14.07^{+16.22}_{-13.83}$ | $2.03^{+0.22}_{-0.21}$ | $1.31^{+0.49}_{-0.37}$ |
| Swift | 2014.09 | 12.16[b] | 1.52[b] | $2.31^{+1.09}_{-0.82}$ |
| Chandra | 2014.12 | $9.49^{+3.25}_{-3.05}$ | $1.52^{+0.06}_{-0.06}$ | $6.96^{+0.61}_{-0.72}$ |
| Swift | 2018.03 | 12.16[b] | 1.52[b] | $2.74^{+1.58}_{-1.56}$ |

**Notes.**
[a] Unabsorbed flux in 2–10 keV.
[b] The values are fixed in the fitting.

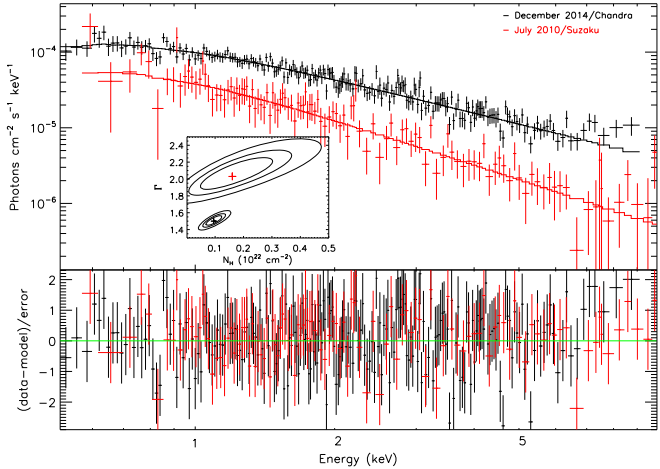

**Figure 3.** Unfolded X-ray spectra of SDSS J155258+273728 with absorbed power-law model fits for *Suzaku* and *Chandra* observations. Inset: contours give the joint 68%, 90%, 99% confidence for the two interesting variables, photon index $\Gamma$, and the intrinsic absorption $N_H$.

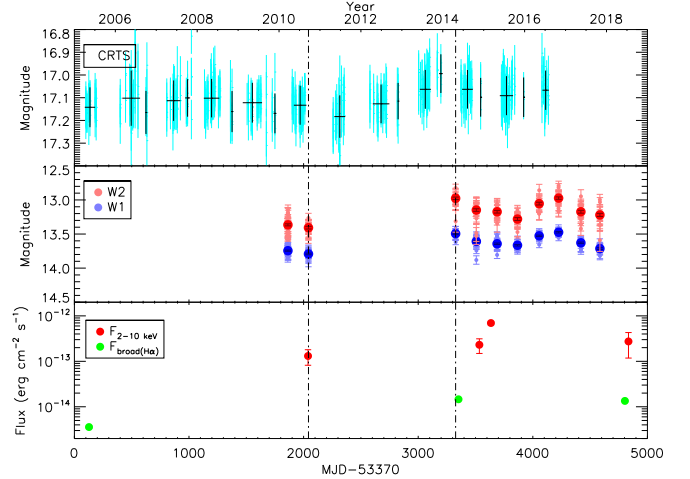

**Figure 4.** Light curves showing the multiwavelength changes observed in SDSS J155258+273728 in the optical *V*-band data from CRTS (top panel), mid-IR *W*1 and *W*2 bands from *WISE* (middle panel), and hard X-ray flux along with broad H$\alpha$ emission (bottom panel). In the top and bottom panels, we also present the median values of each season epoch. The dotted–dashed lines mark the start/end of the significant variation timescale.

show the joint contours of the photon index $\Gamma$ and column density $N_H$. It is clear that the changes of photon index between *Suzaku* and *Chandra* are significant at 99% confidence.

There is more than a factor of 2 brightening in *Chandra* observation with an unabsorbed 2–10 keV flux of $6.96^{+0.61}_{-0.72} \times 10^{-13}$ erg s$^{-1}$ cm$^{-2}$, compared to those in *Suzaku* and *Swift*-2014 with values of $1.31^{+0.49}_{-0.37} \times 10^{-13}$ erg s$^{-1}$ cm$^{-2}$ and $2.31^{+1.09}_{-0.82} \times 10^{-13}$ erg s$^{-1}$ cm$^{-2}$, respectively. The inferred flux of *Swift*-2018 is $2.74^{+1.58}_{-1.56} \times 10^{-13}$ erg s$^{-1}$ cm$^{-2}$, indicating that the X-ray emission of SDSS J155258+273728 fades back to the level of *Swift*-2014.

### 3.1. IR–Optical–X-Ray Photometry and Variability

In Figure 4 we show the X-ray and broad H$\alpha$ line variations of SDSS J155258+273728 on a decade. The optical and mid-IR photometry from CRTS and the *Wide-field Infrared Survey Explorer* survey are also shown. It is clear that accompanied by the significant enhancement of broad H$\alpha$ emission, SDSS J155258+273728 presents increased emission in hard X-ray, optical *V*-band, and mid-IR. The variation timescale of this broadband brightening is ∼4 yr, in which the X-ray varied more than a factor of 2, and CRTS *V*-band and mid-IR *W*1, *W*2 bands varied with amplitudes of 0.18 ± 0.12, 0.32 ± 0.02, and 0.43 ± 0.03, respectively.

### 4. Discussion

#### 4.1. Reddening Measurement from Optical Spectra

The detected moderate absorption in X-ray indicates that there might be obscurations in our line of sight, possibly accounting for the very weak/undetected continuum emission and broad H$\beta$ line in SDSS J155258+273728. We follow the procedures described in Trippe et al. (2010) to measure the upper limit of broad H$\beta$ emission, and to determine if the component should be visible in the spectrum in the absence of reddening. A template representing the intrinsic emitted H$\beta$ line was made with the broad component of Hα at each epoch. We scaled the Hα in width to make up for the velocity width difference at Hβ (Greene & Ho 2005), and divided its flux by 3.0. This template was then added to the spectrum at the position of Hβ. For the SDSS spectrum, the expected intrinsic Hβ is indiscernible against the noise, that is, the broad Hβ line would not be visible even if the broad-line regions were totally unreddened. The result indicates that the undetected broad Hβ line in SDSS spectra may be due to the intrinsic weakness or dust reddened, or both.

For the LAMOST and DBSP spectra, the addition of the template made a visible broad Hβ line, indicating that Hβ emission would be observable at these two epochs in the absence of reddening. We then multiplied the template with progressively smaller scale factors until the line became indistinguishable. The scale factor times the intrinsic expected Hβ flux was then taken as an upper limit to the amount of broad Hβ emission. We show the results in Table 1. The inferred Balmer decrements are steep with values ≲20.6, indicating significant dust reddening to the broad emission-line regions. With the assumption of the standard Galactic reddening curve $R(\lambda)$ of Fitzpatrick & Massa (1999), we calculate the reddening of the broad-line region (BLR) using the following equation:

$$E(B-V) = \frac{-2.5}{R_{H\alpha} - R_{H\beta}} \log \frac{f_{H\alpha}/f_{H\beta}}{3.0}, \quad (1)$$

where the intrinsic flux ratio of the broad Hα and Hβ components is assumed to be $f_{H\alpha}/f_{H\beta} = 3.0$ (Veilleux & Osterbrock 1987; Dong et al. 2008). The low limit of the $E(B-V)_{BLR}$ is 1.63 for these two epochs.

Such extinction might explain the undetected nuclear continuum in the optical spectra. Indeed, if the central nuclear emission is also obscured, the observed flux at the rest wavelength of 5100 Å will be fainter by more than two orders of magnitude than its intrinsic (unabsorbed) value. SDSS J155258+273728 was not detected in the *Galaxy Evolution Explorer* and the *Swift* near- and far-ultraviolet bands. With the empirical $L_{2500\,\text{Å}}$–$L_{2\,\text{keV}}$ relation in AGNs (Lusso et al. 2010), we estimate the intrinsic rest-frame monochromatic luminosity at 2500 Å to be $L_{2500\,\text{Å}} = 2.02 \times 10^{28}$ erg s$^{-1}$ Hz$^{-1}$. Although this value is normal among AGNs, the extinction derived from Balmer decrements makes our object invisible (i.e., below the detection limit) to *GALEX* and the UV bands of *Swift*.

The dust-to-gas ratio in SDSS J155258+273728 is $E(B-V)/N_H \gtrsim 1.15 \times 10^{-21}$ mag cm$^2$, which is larger than the average Galactic value, $1.7 \times 10^{-22}$ mag cm$^{-2}$. This indicates that either the intrinsic Balmer decrements is slightly larger than the assumed value of 3.0 (e.g., Schnorr-Müller et al. 2016), or the dust-to-gas ratio of our object is a higher than the Galactic value.

### 4.2. Black Hole Mass and Accretion Rate

We estimate the black hole mass of SDSS J155258+273728 by virtue of the empirical relation of $M_{BH}$ in AGNs with $L_{H\alpha}$ luminosity and the FWHM of the broad Hα line (Greene & Ho 2005). Here $L_{H\alpha}$ is the unabsorbed (i.e., extinction-corrected) one. The measured black hole mass is similar between SDSS and DBSP epochs (~$2.1 \times 10^8 M_\odot$), which is smaller than that from the LAMOST epoch (~$3.5 \times 10^8 M_\odot$). The consistency in the black hole mass measurement among SDSS and DBSP observations comes from the fact that the inferred Hα line becomes narrower when the system becomes brighter (Table 1). The variations of Hα line followed the "breathing" pattern, in which the broad line width decreased by a factor of 1.42, is almost identical to $(L_{H\alpha,SDSS}/L_{H\alpha,DBSP})^{-0.27} = 1.43$ (Greene & Ho 2005). Note that the Hα profile inferred from LAMOST spectrum deviates from this "breathing" pattern of BLR, likely due to the calibration issues in the LAMOST spectrum. We thus take the black hole mass to be $2.1 \times 10^8 M_\odot$.

We then estimate the bolometric luminosity of SDSS J155258+273728 with 2–10 keV X-ray luminosity using a bolometric correction factor of 15.8 (Vasudevan & Fabian 2007; Cheng et al. 2019; Netzer 2019). SDSS J155258+273728 is fairly faint, with $L_{bol}/L_{Edd} \approx 7 \times 10^{-3}$ in the high state at the *Chandra* epoch, and $\approx 1 \times 10^{-3}$ in the low state at the *Suzaku* epoch. It is well known that low-luminosity AGNs (LLAGNs) whose bolometric luminosity $L_{bol}$ is less than $(0.01-0.02)L_{Edd}$ are distinct to their bright ($L_{bol} \gtrsim (0.01-0.02)L_{Edd}$) cousins (Ho 2008). Both the optical–X-ray spectral index and the X-ray photon index anticorrelate with the Eddington ratio in LLAGNs (e.g., Ho 2008; Yang et al. 2015; Connolly et al. 2016; Xie et al. 2016). The leading theoretical picture of LLAGNs is that the inner region of the accretion flow is no longer a cold optically thick accretion disk (likely the case in bright AGNs), but instead a hot optically thin one (Yuan & Narayan 2014). Note that such change is also observed in BHBs (e.g., Esin et al. 1997; Remillard & McClintock 2006; Done et al. 2007).

### 4.3. Nature of the Variability

As discussed above, the central region of SDSS J155258+273728 is obscured. One may naturally argue that the prominent broad Hα line brightening is due to the passing by of the obscurers along our line of sight. However, the observed optical/mid-IR variability amplitudes and timescales put strong evidence against this natural expectation. First, if the variability is caused by the change in obscuration, then any detectable variability in mid-IR bands implies a much larger variability amplitude in optical. For SDSS J155258+273728 the maximum variation in the $W1$ band is ~0.51 mag, then $\Delta V \sim 11$ mag is required when assuming the extinction model of Fitzpatrick & Massa (1999). This is significantly larger than what we see in Figure 4.

The other evidence against the variable obscuration scenario is related to the timescale. The argument, as widely discussed in the literature (e.g., LaMassa et al. 2015; Sheng et al. 2017), is that the estimated crossing time for an intervening object orbiting outside the BLR on a Keplerian orbit is longer than the variation timescale of broad Balmer lines and mid-IR emission. The crossing time for the obscuring object is measured as $t_{cross} = 0.07[r_{orb}/(\text{lt-day})]^{3/2} M_8^{1/2} \arcsin(r_{src}/r_{orb})$ yr, where $r_{orb}$ is the orbital radius of the foreground object, $M_8$ is the black hole mass in units of $10^8 M_\odot$, and $r_{src}$ is the true size of the obscured region (i.e., the continuum-emitting region or the BLR size; LaMassa et al. 2015). In the measurement we adopt $r_{src}$ as the BLR size, which is estimated from the $R$–$L_{5100}$, $L_{5100}$–$L_{H\alpha}$ relation (Greene & Ho 2005). The size of the obscurer, $r_{orb}$, should be at least comparable to the torus to block the hot dust, from which mid-IR emission at 3.4 and 4.6 μm mainly originated. We take the inner radius of the torus as $r_{orb}$, which was estimated simply from the dust sublimation radius ($R_{sub} = 0.5 L_{46}^{0.5}(1800K/T_{sub})$ pc = 0.098 pc). For SDSS

J155258+273728 the derived value of $t_{cross}$ is ∼15.8 yr, which is longer than the significant variation timescale (∼4 yr).

Thus the most plausible scenario of the dramatic Hα line increments seen in SDSS J155258+273728 is due to the intrinsic variation of the accretion power, which can be clearly illustrated by the X-ray emission. For SDSS J155258+273728, the spectral shape in X-rays varied with luminosity, i.e., the X-ray spectrum is harder (with photon index $\Gamma = 1.52 \pm 0.06$) at the bright *Chandra* epoch, than that at the faint *Suzaku* epoch ($\Gamma = 2.03^{+0.22}_{-0.21}$). Interestingly, such "harder when brighter" spectral behavior in X-rays is observed in both BHBs in their hard states and LLAGNs (e.g., Kalemci et al. 2005; Wu & Gu 2008; Gu & Cao 2009; Sobolewska et al. 2011; Younes et al. 2011; Yang et al. 2015; Connolly et al. 2016), all of which are below a critical luminosity of around a few percent of the Eddington luminosity. Above this critical luminosity, i.e., for BHBs in intermediate and soft states and bright AGNs, the X-ray spectrum softens as the flux increases ("softer when brighter"; e.g., Zdziarski et al. 2003; Wang et al. 2004; Risaliti et al. 2009; Sobolewska & Papadakis 2009; Trakhtenbrot et al. 2017). Unlike BHBs, most AGNs only stay in one branch because of small variation in X-ray luminosity of individual sources; the only exception to our knowledge is the LLAGN NGC 7213 (Xie et al. 2016).

In this work, we argue that the change in accretion power (or, equivalently, accretion rate) is not sufficient to make AGNs change their look, a state transition is further required. More clearly, we propose the appearance of the broad Hα component in the bright AGN regime and disappearance of the broad Hα component in the LLAGN regime. We note that similar scenarios have been proposed in the literature (Noda & Done 2018; Dexter & Begelman 2019a; Ruan et al. 2019), although the on–off timescale in CL-AGNs remains a challenging task in this model (Dexter & Begelman 2019a). Based on broadband continuum spectral modeling Noda & Done (2018) suggest that Mrk 1018 underwent a soft-to-hard state transition as it faded from Seyfert 1 to 1.9. Ruan et al. (2019) claimed that the correlation between the UV-to-X-ray spectral index and Eddington ratio in CL-AGNs is similar to the spectral evolution of BHBs. These results are intriguing. However, the application of this method is limited. The broadband continuum spectral modeling can only be applied to a limited number of sources, and the UV-to-X-ray spectra index is difficult to constrain at a low-luminosity state, where emission from host galaxy dominates.

To statistically study the accretion physics of CL-AGNs, we compiled from literature the X-ray data (Longinotti et al. 2007; Denney et al. 2014; LaMassa et al. 2015; Noda & Done 2018; Parker et al. 2019). As shown in Figure 5, we examined the relation between the hardness ratio, $F_{soft}/(F_{soft}+F_{hard})$, and the hard X-ray luminosity (in Eddington units), $L_{X,\,2-10\,keV}/L_{Edd}$. Here $F_{soft}$ and $F_{hard}$ are the absorption-corrected flux in the soft (0.3–2 keV) and hard (2–10 keV) bands, respectively. As clearly shown in Figure 5, we observe in CL-AGNs a "*V*"-shaped relationship between the hardness ratio and $L_{X,\,2-10\,keV}/L_{Edd}$, where the turnover locates at $L_{X,\,2-10\,keV}/L_{Edd} \sim 10^{-3}$ (note that the exact value varies among individual sources). For comparison, we also plot the fits to the relation from previous samples of LLAGNs (Constantin et al. 2009) and of bright AGNs (Risaliti et al. 2009), with the conversion of $\Gamma$ to hardness ratio and an assumed bolometric correction factor of $L_{bol}/L_x = 16$. Admittedly, more data are needed to examine the spectral behavior of CL-AGNs at the $L_{X,\,2-10\,keV}/L_{Edd} \lesssim 10^{-3}$ regime (Liu et al. 2019). Additionally from Figure 5 we learn that as the optical spectral state evolves from type 1 to type 1.9/2, the X-ray spectral shape transits from soft to hard, or more aggressively from the positive branch to the negative branch (of the $F_{soft}/(F_{soft}+F_{hard})$–$L_{X,\,2-10\,keV}/L_{Edd}$ correlation).

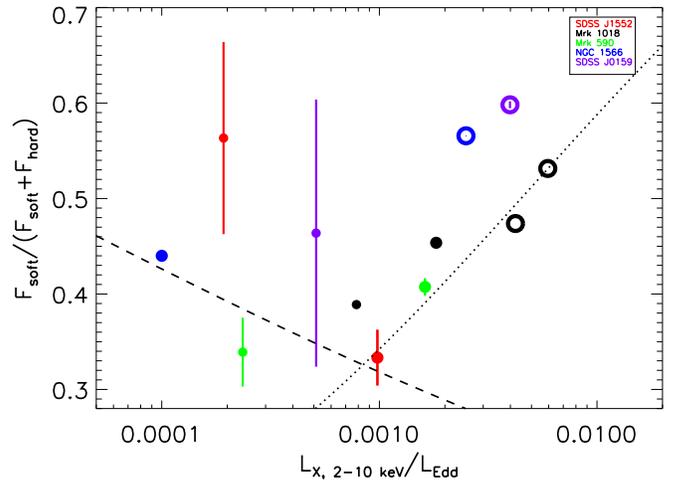

**Figure 5.** Variation of hardness ratio with the Eddington-scaled X-ray luminosity for changing-look AGNs. The $F_{soft}$ and $F_{hard}$ are the fluxes in 0.3–2 keV and 2–10 keV, respectively. Symbols with a larger size represent the states with enhanced broad Balmer emission lines. Filled circles represent the optical spectral states of type 1.9/2, and open circles represent the type 1 state. The dashed line is the fit from Constantin et al. (2009), and the dotted line is the fit from Risaliti et al. (2009). The data for Mrk 1018, Mrk 590, and NGC 1566 and SDSS J015957.64+003310 are from the literature (Longinotti et al. 2007; Denney et al. 2014; LaMassa et al. 2015; Noda & Done 2018; Parker et al. 2019).

Our results provide direct evidence that the change in the optical spectral type of CL-AGNs associates with a change in the evolution trend of the X-ray spectral shape. Theoretically, the change in the evolution of the X-ray spectral shape is normally interpreted as a consequence of the change in accretion mode, i.e., the region proximity to the black hole varies between a hot accretion flow and a two-phase accretion flow (either numerous cold clumps embedded in a hot flow, or at even higher luminosities in a cold disk sandwiched by hot coronas; Gardner & Done 2013; Qiao & Liu 2013; Yang et al. 2015; Xie et al. 2016). Under this well-accepted model, the hot accretion flow represents the hard state of BHBs and LLAGNs, the two-phase accretion represents the intermediate state of BHBs, and the disk–corona configuration is for the soft state of BHBs and bright AGNs (Esin et al. 1997; Remillard & McClintock 2006; Done et al. 2007; Yang et al. 2015). The similarity between our result and the normal evolution of AGNs may suggest that the CL-AGNs represent a unique evolutionary stage that undergoes the accretion state transition (see also Noda & Done 2018; Ruan et al. 2019; Dexter & Begelman 2019a).

## 5. Conclusion

We present multiband spectra and variability studies of the changing-look AGN SDSS J155258+273728. We find:

1. A prominent broad Hα line was still detected in the recent 2018–DBSP spectra. The steep Balmer decrements indicate our line of sight might be obscured, which possibly accounts for the very weak/undetected continuum nuclear emission and broad H$\beta$ line.

2. The object presents significant variations in X-ray, optical, and mid-IR bands. Timescales and amplitudes in these multiband variations provide strong evidence against the variable obscuration scenario, indicating intrinsic emission varied in this CL-AGN.

3. The X-ray spectral shape varied with the luminosity in a way of "harder when brighter." In compiled studies of CL-AGNs we find that the hardness ratio increases with luminosity, while, at accretion rates below $L_{X,2-10\,keV}/L_{Edd} \sim 10^{-3}$, there seems to be a turnover of the relation. There is also evidence that changes in the optical spectral type associate with the changes in the X-ray spectral shape transition. These similarities of the X-ray spectral evolution to outbursting X-ray binaries support the accretion state transition scenarios of CL-AGNs.

Future studies of more CL-AGNs with high-S/N multiband data will provide more clues about the accretion physics happening at the region proximity to central black holes.

We acknowledge the anonymous referee for valuable comments that helped to improve the Letter. This work is supported by the National Science Foundation of China (NSFC) grant U1731103, 11833007, and 11833001. L-M Dou acknowledge supports from NSFC U1731103, 11833007, and 11833001. F. G. Xie was supported in part by National Program on Key R&D Project of China (NPKRDPC) 2016YFA0400804 and NSFC 11873074. X.-B. Wu thanks the supports by the NPKRDPC 2016YFA0400703 and the NSFC 11533001, 11721303. X.S. acknowledges the support by NSFC 11822301. We thank the Swift Acting PI, Brad Cenko, for approving our ToO request, and the Swift observer team. This work uses data obtained through the Telescope Access Program (TAP), which has been funded by the National Astronomical Observatories, Chinese Academy of Sciences, and the Special Fund for Astronomy from the Ministry of Finance. Observations obtained with the Hale Telescope at Palomar Observatory were obtained as part of an agreement between the National Astronomical Observatories, Chinese Academy of Sciences, and the California Institute of Technology. We acknowledge the use of LAMOST data. The Large Sky Area Multi-Object Fiber Spectroscopic Tele-scope (LAMOST, also named Guoshoujing Telescope) is a National Major Scientific Project built by the Chinese Academy of Sciences. This publication makes use of data products from the Widefield Infrared Survey Explorer, which is a joint project of the University of California, Los Angeles, and the Jet Propulsion Laboratory/California Institute of Technology, funded by the National Aeronautics and Space Administration.

### ORCID iDs

Yanli Ai https://orcid.org/0000-0001-9312-4640
Liming Dou https://orcid.org/0000-0002-4757-8622
Fu-Guo Xie https://orcid.org/0000-0001-9969-2091
Su Yao https://orcid.org/0000-0002-9728-1552
Xue-Bing Wu https://orcid.org/0000-0002-7350-6913
Ning Jiang https://orcid.org/0000-0002-7152-3621